\documentclass[12pt,a4paper,titlepage]{article}
\usepackage{amsmath,color}
\usepackage{a4wide}

\renewcommand{\eqref}[1]{(\ref{#1})}

\usepackage{bm}
\usepackage{here}

\definecolor{hyprf}{cmyk}{1,0.5,0,0}

\usepackage[utf8]{inputenc}
\usepackage[top=50pt,bottom=60pt,left=68pt,right=66pt]{geometry}
\usepackage{amsmath}
\usepackage{booktabs}
\usepackage{amssymb}
\usepackage[title]{appendix}
\usepackage{mathrsfs}
\usepackage{float}
\usepackage{xcolor}
\usepackage{multirow}
\usepackage{graphicx,caption,subcaption}
\usepackage{comment}
\usepackage[space]{grffile}

\def\un#1{\relax\ifmmode\@@underline#1\else
        $\@@underline{\hbox{#1}}$\relax\fi}

\let\du=\du                     

\def\a{\alpha}

\def\m{\mu}

\def\z{\zeta}
\def\D{\Delta}

\def\L{\Lambda}

\def\bo{{\raise-.3ex\hbox{\large$\Box$}}}               
\def\pa{\partial}                                       
\def\TH{{\raise.2ex\hbox{$\displaystyle \bigodot$}\mskip-4.7mu \llap H \;}}
\def\face{{\raise.2ex\hbox{$\displaystyle \bigodot$}\mskip-2.2mu \llap {$\ddot
        \smile$}}}                                      

\def\abs#1{\left| #1\right|}                    
\def\leftrightarrowfill{$\mathsurround=0pt \mathord\leftarrow \mkern-6mu
        \cleaders\hbox{$\mkern-2mu \mathord- \mkern-2mu$}\hfill
        \mkern-6mu \mathord\rightarrow$}
\def\dvec#1{\vbox{\ialign{##\crcr
        \leftrightarrowfill\crcr\noalign{\kern-1pt\nointerlineskip}
        $\hfil\displaystyle{#1}\hfil$\crcr}}}           

\def\frac#1#2{{\textstyle{#1\over\vphantom2\smash{\raise.20ex
        \hbox{$\scriptstyle{#2}$}}}}}                   
\def\sfrac#1#2{{\vphantom1\smash{\lower.5ex\hbox{\small$#1$}}\over
        \vphantom1\smash{\raise.4ex\hbox{\small$#2$}}}} 
\def\bfrac#1#2{{\vphantom1\smash{\lower.5ex\hbox{$#1$}}\over
        \vphantom1\smash{\raise.3ex\hbox{$#2$}}}}       
\def\afrac#1#2{{\vphantom1\smash{\lower.5ex\hbox{$#1$}}\over#2}}    

\def\[{\lfloor{\hskip 0.35pt}\!\!\!\lceil}
\def\]{\rfloor{\hskip 0.35pt}\!\!\!\rceil}

\def\du#1#2{_{#1}{}^{#2}}

\def\un{\underline}
\def\fracmm#1#2{{{#1}\over{#2}}}

\def\low#1{{\raise -3pt\hbox{${\hskip 0.75pt}\!_{#1}$}}}

\newskip\humongous \humongous=0pt plus 1000pt minus 1000pt

\newif\ifdtup

\def\({\left(}
\def\){\right)}

\def\beq{\begin{equation}}
\def\eeq{\end{equation}}
\def\bea{\begin{eqnarray}}
\def\eea{\end{eqnarray}}

\newcommand{\be}{\begin{equation}}
\newcommand{\ee}{\end{equation}}
\newcommand{\nbe}{\begin{equation*}}
\newcommand{\nee}{\end{equation*}}

\newcommand{\lb}{\label}

\begin{document}
\renewcommand{\arraystretch}{1.3}

\thispagestyle{empty}

\noindent {\hbox to\hsize{
\vbox{\noindent January 2023 \hfill IPMU22-0049 }}
\noindent  $~$ revised version \hfill }

\noindent
\vskip2.0cm
\begin{center}

{\Large\bf Pole inflation and primordial black holes \\ formation in Starobinsky-like supergravity}

\vglue.3in

Shuntaro Aoki~${}^{a}$, Ryotaro Ishikawa~${}^{b}$, and Sergei V. Ketov~${}^{b,c,d}$
\vglue.3in

${}^a$~Department of Physics, Chung-Ang University, Seoul 06974, South Korea\\
${}^b$~Department of Physics, Tokyo Metropolitan University\\
1-1 Minami-ohsawa, Hachioji-shi, Tokyo 192-0397, Japan \\
${}^c$~Research School of High-Energy Physics, Tomsk Polytechnic University\\
2a Lenin Avenue, Tomsk 634028, Russian Federation\\
${}^d$~Kavli Institute for the Physics and Mathematics of the Universe (WPI)
\\The University of Tokyo Institutes for Advanced Study, Kashiwa 277-8583, Japan\\
\vglue.1in

shuntaro@cau.ac.kr, ishikawa-ryotaro@ed.tmu.ac.jp, ketov@tmu.ac.jp
\end{center}

\vglue.3in

\begin{center}
{\Large\bf Abstract}
\end{center}
\vglue.2in
We extend the Cecotti-Kallosh model of Starobinsky inflation in supergravity by adding a holomorphic function to the 
superpotential in order to generate a large peak in the power spectrum of scalar (curvature) perturbations. In our approach, 
the singular non-canonical kinetic terms are largely responsible for inflation (as an attractor solution), whereas the superpotential  is engineered to generate a production of primordial black holes. We study the cases with (i) a linear holomorphic function, (ii) a quadratic holomorphic function, and (iii) an exponential holomorphic function, as regards the dependence of inflation and primordial black holes production upon parameters of those functions and initial conditions,
as well as verify viability of inflation with our superpotentials. We find that an efficient production of primordial black holes consistent with CMB measurements is only possible  in the second (ii) case. We calculate the masses of the produced
primordial black holes and find that they are below the Hawking (black hole) evaporation limit, so that they cannot be part of
the current dark matter in our Universe. 

\newpage

\section{Introduction}

Cosmological inflation in the very early Universe is well motivated both theoretically and experimentally. It solves the old problems of the 
standard (Friedman) cosmology, proposes the origin of the large scale structure, and gives correct predictions about fluctuations and anisotropy
of the cosmic microwave background (CMB) radiation.  

The inflationary paradigm is not yet an established theory because the scale of inflation, its driver (inflaton) and the origin of the inflaton scalar potential are unknown. There are many inflationary models in the literature so that there is a need for their discrimination. A necessary requirement of consistency with CMB measurements still leaves a plenty of viable inflationary models because the CMB window is very small, being
limited to the scales (wavenumbers) $k$ between $10^{-4}$ ${\rm Mpc}^{-1}$ and $5\cdot 10^{-1}$ ${\rm Mpc}^{-1}$. Amongst the additional theoretical requirements can be demands for an 
attractor solution for inflation, the ultra-violet cutoff well beyond the Hubble value during inflation, a minimal number of interactions and parameters,
a clear physical origin of inflaton, additional fundamental symmetries, etc. It is, therefore, desirable to separate the well motivated features of viable inflationary models from environmental (or model-dependent) features specific to particular models of inflation, as well as identify the leading candidate by applying the former conditions. 

For this purpose, in this paper we employ the framework of {\it pole inflation}  that allows one to essentially describe inflation by a non-canonical kinetic term  having a pole, with a related unification of inflationary models into the universality classes. This way of reasoning leads to the  cosmological $\alpha$-attractors  \cite{Galante:2014ifa,Terada:2016nqg,Kallosh:2022feu}. Among the inflationary $\alpha$-attractors, the famous Starobinsky model (1980) of inflation \cite{Starobinsky:1980te} still occupies the leading position because (i) it employs only gravitational interactions, (ii) Starobinsky's inflaton (scalaron) is a physical excitation of the higher-derivative gravity and can be interpreted as the Nambu-Goldstone boson associated with spontaneous breaking of scale invariance in the no-scale and no-ghost $R^2$-gravity, (iii) the ultra-violet cutoff in the Starobinsky model is given by the (reduced) Planck mass $M_{\rm Pl}$ that is beyond the Hubble function $H$ during inflation by the five orders of the magnitude, (iv) the predictions of the Starobinsky model for CMB are in excellent agreement with current (2021) measurements of CMB  \cite{BICEP:2021xfz,Tristram:2021tvh}, (v) the Starobinsky model has only one parameter given by the inflaton mass $m$ that is fixed by the CMB amplitude as $m={\cal O}(10^{-5})M_{\rm Pl}$, so that the Starobinsky model of inflation has no free parameters at all, (vi) the Starobisky inflation offers the universal mechanism for reheating after inflation, see e.g., Refs.~\cite{Ketov:2010qz,Ketov:2012yz,Ketov:2019toi} for a review of all these features.

The inflationary scale $H\sim {\cal O}(10^{14})$ GeV in the Starobinsky model implies the necessity to include new physics well beyond the electro-weak scale because the value of $H$ is not far from the Grand Unification scale where gravitational interactions of elementary particles can no longer be ignored. The appropriate framework in the theoretical high energy physics at those scales is given by {\it supergravity} \cite{Ellis:2013xoa,Ketov:2012yz}.  In this paper, we employ the Starobinsky inflation in the supergravity framework. 

The use of pole inflation allows us to describe slow-roll inflation mainly by the kinetic terms, with the scalar potential being largely undetermined. Then we can use freedom in our choice of the scalar potential for engineering a formation of primordial black holes (PBH) during or after inflation. In supergravity theory, the kinetic terms are described by a K\"ahler potential, and the scalar potential is governed by a superpotential also. We use the supergravity description of the Starobinsky inflation in the minimal form of the Cecotti-Kallosh model \cite{Cecotti:2014ipa} but modify their superpotential by a holomorphic function towards an inclusion of PBH production. It leads to the two-field inflation whose consistency has to be checked again because the modified superpotentials can destroy viable inflation.  Multi-field inflation and its applications in supergravity for PBH production were investigated in the different models in Refs.~\cite{Gundhi:2020kzm,Aldabergenov:2020bpt,Ketov:2021fww,Geller:2022nkr,Aldabergenov:2022rfc}.

Our paper is organized as follows. Section 2 is our setup where we introduce our model, the parameterizations of the scalar fields used, and demonstrate consistency with the earlier results. The main body of our paper is devoted to the simplest Ans\"atze for the new functions $g$ and
$\tilde{g}$ defined in Sec.~2. First, we try linear functions for them  and investigate their parameter space in Sec.~3, as regards the impact for
double inflation and PBH formation. Next, in Section 4, we investigate the case of a quadratic $g$-function that turns out to be our main case. In Section 5 we derive the inflationary trajectory, the slow roll parameters and the CMB observables in the case of the quadratic $g$-function because other choices do not work. In Section 6 we study the dependence of the observables upon initial conditions. In Section 7 we compute the power spectrum of scalar perturbations and the masses of the generated PBH. Our conclusion is Section 8. We also study some other possible cases, including the exponential functions for $g$ and $\tilde{g}$, and find that they do not work for inflation and PBH production, see Appendices A and B  too.

We do not provide a historical overview or an introduction to cosmological inflation and primordial black holes, skip some basic equations because they can be easily found in many publications, see e.g., Refs.~\cite{Ketov:2012yz,Gundhi:2020kzm,Ketov:2021fww}, and confine our paper to original new results derived either analytically or numerically,  with a minimum of relevant references. We use the natural units with $M_{\rm Pl}=1$ throughout the paper.

\section{Setup}

The basic idea of pole inflation \cite{Galante:2014ifa,Terada:2016nqg,Kallosh:2022feu} can be illustrated on the simplest example of the single-field E-models of inflation, also known as the cosmological $\a$-attractors in the literature, by starting from the Lagrangian 
\be \lb{poleinf}
{\cal L}[y]= -\fracmm{3\a}{4y^2}\pa^{\m}y \pa_{\m}y - f^2(y)
\ee
of the real field $y(x^{\m})$ having the singular kinetic term (with a pole of the 2nd order at $y=0$), the parameter $\a>0$ and the potential $V(y)=f^2(y)$ in terms of a real non-constant function $f(y)$ that is analytic at the origin $y=0$ with $f(0)\neq 0$. A non-linear field redefinition 
\be \lb{y}
y= \exp\left(-\sqrt{\fracmm{2}{3\a}}\varphi\right)
\ee
 yields the canonical kinetic term of $\varphi$ with the scalar potential $V(\varphi)=f^2(y(\varphi))$ that can be expanded in power series with respect to $y$ that is supposed to be small during slow-roll inflation \cite{Ketov:2021fww,Ivanov:2021chn}. The linear term (in $y$) of that expansion contributes to the leading terms (with respect to the inverse powers of the e-folds number $N_e$) in the cosmological observables known as the tilt $n_s$ of scalar perturbations and the tensor-to-scalar ratio $r$, so that the rest of the expansion can be chosen at will or for other purposes.

The pole inflation makes manifest the universality classes of the cosmological $\a$-attractors, which are parametrized by 
$\a$ \cite{Kallosh:2013hoa,Ketov:2019toi}. Their predictions for the CMB spectrum tilts,
\be \lb{alphap}
n_s=1 - \fracmm{2}{N_e}~,\quad r= \fracmm{12\a}{N_e^2}~~,
\ee
comfortably fit observations \cite{BICEP:2021xfz,Tristram:2021tvh} for the parameter $\a$ values around one and $N_e=55\pm 10$. 
The Starobinsky inflation \cite{Starobinsky:1980te} appears in the case of $\a=1$, while there exist the simple dual version of the model (\ref{poleinf})  known as the modified $(R+R^2)$ gravity, see e.g., Refs.~\cite{Ketov:2021fww,Ivanov:2021chn} for a recent review. In this paper, we confine ourselves to $\a=1$ and dub our modifications Starobinsky-like accordingly.

Our supergravity extension of the Starobinsky model of inflation in this paper is described by the K\"ahler potential and the superpotential (cf. Refs.~\cite{Kawasaki:2000yn,Ellis:2013xoa,Cecotti:2014ipa})
as follows:
\begin{align}
&K=-3\ln \left(T+\bar{T}-|C|^2+\zeta\fracmm{|C|^4}{T+\bar{T}}\right),\label{K}\\
&W=MC(T-1)+g(T),\label{W}
\end{align}
in terms of the inflaton superfield $T$ and the goldstino superfield $C$. The inflaton is the scalar field component of the superfield $T$. Inflation spontaneously breaks supersymmetry, which leads to the Nambu-Goldstone fermion called goldstino that is the field component of the superfield $C$. We use the same notation for superfields and their first
field components. The parameter $M$ is proportional to the mass $m_{\rm inf.}$ of Starobinsky's scalaron (inflaton), $m_{\rm inf.}\sim 10^{-5}M_{\rm Pl}$. The term with the real coefficient $\z$ inside the logarithm is needed for stabilization of the inflationary trajectory in the scalar field space with a $T$-independent mass at $C=0$ ~\cite{Cecotti:2014ipa}. When 
$\zeta=g(T)=0$, our model (\ref{W}) reduces to the model in Refs.~\cite{Cecotti:1987sa,Gates:2009hu}.  Unlike those references, we have added a new analytic function $g(T)$ to the superpotential (\ref{W}) that will be needed for PBH production. 

 In terms of the leading (scalar) field components, Eqs.~(\ref{K}) and (\ref{W}) give the Lagrangian 
\begin{align} \lb{tk}
&\mathcal{L}=-\fracmm{3}{(T+\bar{T})^{2}} \partial_{\mu} T \partial^{\mu} \bar{T}-V,\\
&V=\frac{1}{3} \fracmm{M^2|T-1|^2}{(T+\bar{T})^{2}}+\frac{1}{3} \fracmm{1}{(T+\bar{T})}\left|\fracmm{dg}{dT}\right|^{2}-\fracmm{\frac{dg}{dT} \bar{g}+\fracmm{d\bar{g}}{d\bar{T}} g}{(T+\bar{T})^{2}}~~, \lb{tp}
\end{align}
where we have set $C=0$, and the bars denote complex conjugation. 

There is a pole in the kinetic term (\ref{tk})  at Re$~T=0$. However, unlike Eq.~(\ref{poleinf}), there are two real  scalars, while their kinetic terms are field-dependent, i.e. they form a non-linear sigma-model (NLSM) \cite{Ketov:2000dy}. 

The half-(complex)-plane variable $T$ can be exchanged to the (Poincar\'e disk) variable $Z$ via the Cayley
(holomorphic) transformation accompanied by a similar change of variables from $C$ to $S$ as follows:
\begin{align} \lb{cayley}
T=\fracmm{1+Z}{1-Z}, \quad C= \fracmm{\sqrt{2}S}{1-Z}~~,
\end{align}
with the inverse transformation
\begin{align} \lb{icayley}
Z=\fracmm{T-1}{1+T}, \quad S= \fracmm{\sqrt{2}C}{1+T}~~.
\end{align}

Equations (\ref{K}) and (\ref{W}) can then be rewritten to
\begin{align}
&K=-3\ln \left(1-|Z|^2-|S|^2+\zeta\fracmm{|S|^4}{1-|Z|^2}\right)+3\ln \fracmm{|1-Z|^2}{2}~,\\
&W=2\sqrt{2}\fracmm{MZS}{(1-Z)^2}+g(Z)~.
\end{align}
Next, after a K\"ahler transformation, $K\rightarrow K+\L+\bar{\L}$ and $W\rightarrow e^{-\L}W$, by
choosing $\L=-3\log (1-Z)/\sqrt{2}$, we obtain
\begin{align}
&K=-3\ln \left(1-|Z|^2-|S|^2+\zeta\fracmm{|S|^4}{1-|Z|^2}\right)~,\label{K2}\\
&W=MZ(1-Z)S+\tilde{g}(Z)~,\label{W2}
\end{align}
where we have introduced  
\be \lb{tildegdef}
\tilde{g}(Z)=\fracmm{(1-Z)^3g(Z)}{2\sqrt{2}}~~.
\ee

The scalar Lagrangian at $S=0$ reads
\begin{align}
&\mathcal{L}=- \fracmm{3}{\left(1-|Z|^{2}\right)^{2}} \partial_{\mu} Z \partial^{\mu} \bar{Z}-V,
\end{align}
where the scalar potential is given by
\begin{align} \lb{pot}
\nonumber V=&\ \frac{M^2}{3} \fracmm{|Z|^2\left|1-Z\right|^{2}}{\left(1-|Z|^{2}\right)^{2}}+\frac{1}{3} \fracmm{1}{(1-|Z|^{2})}\left|\fracmm{d\tilde{g}}{dZ}+\fracmm{3\bar{Z}}{(1-|Z|^{2})}\tilde{g}\right|^{2}-\fracmm{3}{\left(1-|Z|^{2}\right)^{3}}|\tilde{g}|^2 \\
=&\ \frac{M^2}{3} \fracmm{|Z|^2\left|1-Z\right|^{2}}{\left(1-|Z|^{2}\right)^{2}}+\frac{1}{24} \fracmm{|1-Z|^{6}}{1-|Z|^{2}}\left|\fracmm{dg}{dZ}\right|^{2}-\frac{1}{8} \fracmm{|1-Z|^{4}}{\left(1-|Z|^{2}\right)^{2}}\left[(1-Z)^{2} \fracmm{dg}{dZ} \bar{g}+(1-\bar{Z})^{2} \fracmm{d \bar{g}}{d \bar{Z}} g\right].
\end{align}

There is no fundamental reason to prefer any of the two parameterizations ($T$ or $Z$). The models related by the map (\ref{tildegdef}) are equivalent. However, Eq.~(\ref{tildegdef})  does not map a polynomial to a polynomial of the same order. Being interested in minimizing the number of the parameters, we restrict ourselves to the lowest order polynomial 
superpotentials in both parameterizations.

It is convenient to parametrize 
\be \lb{paraz} Z=re^{i\theta}={\rm{tanh}}  \fracmm{\varphi}{\sqrt{6}}\,e^{i\theta}~~.
\ee
Then the kinetic term of $Z$ can be rewritten to
\begin{align}
\mathcal{L}_{\rm{kin}}=-\frac{1}{2} (\partial_{\mu}\varphi)^2-\frac{3}{4}{\rm{sinh}}^2 \fracmm{2\varphi}{\sqrt{6}}(\partial_{\mu}\theta)^2,    
\end{align}
where the field $\varphi$ can be identified with the canonical inflaton (scalaron), and $\theta$ is the additional physical scalar field (sinflaton). It is not possible to canonically normalize both fields because the field space (NLSM) curvature is  non-vanishing. We get the NLSM  metric  
\begin{align}
G_{ab}= \left(\begin{array}{cc}
1 & 0 \\
0 & \frac{3}{2}{\rm{sinh}}^2 \fracmm{2\varphi}{\sqrt{6}}
\end{array}\right),    
\end{align}
with $a,b=(1,2)=(\varphi, \theta)$. The non-vanishing Christoffel symbols are
\begin{align}
\Gamma_{\theta\theta}^{\varphi}=-\fracmm{3}{2\sqrt{6}} {\rm{sinh}} \fracmm{4\varphi}{\sqrt{6}}~, \quad \Gamma_{\theta\varphi}^{\theta}= \Gamma_{\varphi\theta}^{\theta}=\fracmm{2}{\sqrt{6}} {\rm{coth}} \fracmm{2\varphi}{\sqrt{6}}~, 
\end{align}
which give rise to the NLSM Ricci scalar 
\begin{align}
\mathcal{R}=-\fracmm{4}{3}~.   
\end{align}
Therefore, the NLSM has the constant negative curvature in the field space, i.e. a hyperbolic geometry.

In the case of $g=0$, the scalar potential (\ref{pot})  reduces to
\begin{align}
V=\fracmm{M^2}{12} {\rm{sinh}}^2 \fracmm{2\varphi}{\sqrt{6}}\left(1+{\rm{tanh}}^2  \fracmm{\varphi}{\sqrt{6}}-2{\rm{tanh}}  \fracmm{\varphi}{\sqrt{6}}{\rm{cos}}\theta \right).
\end{align}

Let us demonstrate that $\theta$ is stabilized with a heavy mass during slow-roll inflation. The second derivative of the potential, $d^2V/d\theta^2$, at $\theta=0$ is given by
\begin{align} \lb{sd}
\left. \fracmm{d^2V}{d\theta^2}\right|_{\theta=0}=  \fracmm{M^2}{6} {\rm{sinh}}^2 \fracmm{2\varphi}{\sqrt{6}}{\rm{tanh}}  \fracmm{\varphi}{\sqrt{6}}~~.  
\end{align}
When $\varphi$ is approximately constant, the canonical normalization of the $\theta$-field is 
\begin{align}\lb{ctheta}
\tilde{\theta}=\sqrt{\fracmm{3}{2}}{\rm{sinh}} \fracmm{2\varphi}{\sqrt{6}} \cdot \theta~,  
\end{align}
so that we can read off the effective mass of $\tilde{\theta}$ from Eqs.~(\ref{sd}) and (\ref{ctheta}) as
\begin{align}
m^2_{\tilde{\theta}}= \fracmm{M^2}{9} {\rm{tanh}}  \fracmm{\varphi}{\sqrt{6}} \simeq 4H^2~,
\end{align}
where we have used the Friedman equation. Since the $\tilde{\theta}$-field has the mass beyond the Hubble scale $H$ during the first (Starobinsky) stage of inflation, we can safely ignore isocurvature perturbations in the $\tilde{\theta}$-direction (but not later: a tachyonic  instability in the $\theta$-direction will be needed for PBH production after Starobinsky inflation, see next Sections).

\begin{figure}[h]
  \begin{tabular}{cc}
    \begin{minipage}[t]{0.45\hsize}
      \centering
      \includegraphics[scale=0.7]{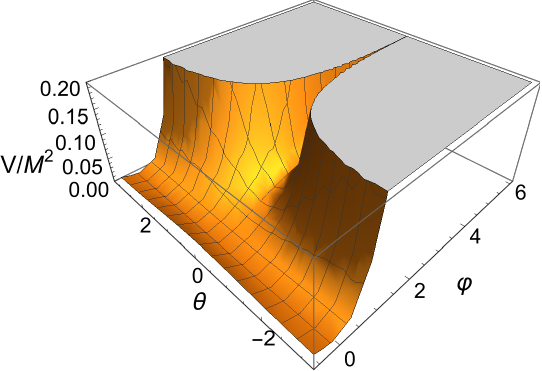}
      \label{v0}
    \end{minipage} &
    \begin{minipage}[t]{0.45\hsize}
      \centering
      \includegraphics[scale=0.7]{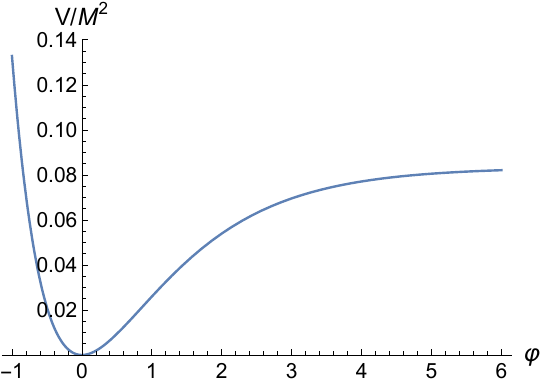}
      \label{v0p}
    \end{minipage}
  \end{tabular}
  \caption{The profile of the potential for $g=0$ (left) and its slice at $\theta=0$ (right).}
  \label{pot0}
\end{figure}

Having stabilized the $\theta$-field during the first stage of inflation, we get the effective single-field potential for inflation as follows:
\be \lb{pstar}
V_{\rm{eff}.}=\fracmm{M^2}{3}\fracmm{{\rm{tanh}}^2  \fracmm{\varphi}{\sqrt{6}}}{(1+{\rm{tanh}}  \fracmm{\varphi}{\sqrt{6}})^2}=\fracmm{M^2}{12}\left(1-e^{-\sqrt{\fracmm{2}{3}}\varphi}\right)^2   
=\fracmm{4m_{\rm inf.}^2}{3}\left(1-e^{-\sqrt{\fracmm{2}{3}}\varphi}\right)^2   
\ee
that coincides with the standard potential of the Starobinsky model. The shape of the
two-field potential and its slice at $\theta=0$ are given in Fig.~\ref{pot0}.

When the function $g\neq 0$, both scalars should be taken into account, and single-field inflation becomes  two-field inflation.  The corresponding equations of motion in the Friedman universe read
\begin{align}
  &0=\ddot{\varphi}+3H\dot{\varphi}-\sqrt{\frac{3}{8}}\sinh\left(\sqrt{\frac{8}{3}}\varphi\right)\dot{\theta}^2+\partial_\varphi V~~,\label{eq1}\\
  &0=\ddot{\theta}+3H\dot{\theta}+2\sqrt{\frac{2}{3}}{\mathrm {coth}}{\left(\sqrt{\frac{2}{3}}\varphi\right)}\dot{\varphi}\dot{\theta}+\frac{2}{3}\mathrm{csch}^2\left(\sqrt{\frac{2}{3}}\varphi\right)\partial_\theta V~~,\label{eq2}\\
  &0=\frac{1}{2}\dot{\varphi}^2+\frac{3}{4}\sinh^2\left(\sqrt{\frac{2}{3}}\varphi\right)\dot{\theta}^2-3H^2+V~~,\label{eq3}
\end{align}
with the potential 
\begin{align}
  V=&\fracmm{M^2}{3}\fracmm{r^2(1-2r\cos\theta+r^2)}{(1-r^2)^2}+\fracmm{1}{24}\fracmm{(1-2r\cos\theta+r^2)^3}{1-r^2}\left|\fracmm{\mathrm{d}g}{\mathrm{d}Z}\right|^2\nonumber\\
    &-\fracmm{1}{8}\fracmm{(1-2r\cos\theta+r^2)^2}{(1-r^2)^2}\left[(1-re^{i\theta})^2\fracmm{\mathrm{d}g}{\mathrm{d}Z}\bar{g}
    +(1-re^{-i\theta})^2\fracmm{\mathrm{d}\bar{g}}{\mathrm{d}\bar{Z}}g\right]~~.
\end{align}

\section{Adding a linear superpotential}

Having the $T$- and $Z$-parameterizations on equal footing, we have to choose either a $g$-function or
a $\tilde{g}$-function, respectively. The simplest choices are given by rational functions and exponentials. In Appendices A and B we demonstrate that the exponentials are ruled out because they destabilize inflation and do not lead to PBH production. Hence, we consider only polynomial functions $g(Z)$ and $\tilde{g}(Z)$ in the main text.  It follows from Eq.~(\ref{tildeg}) that a polynomial function $g(Z)$ leads to a polynomial function $\tilde{g}(Z)$ but not vice versa in general.

\subsection{A linear $g$-function}

Let us use a linear function as the first trial,
\be \lb{ling}
g(Z)=M(g_0+g_1Z)
\ee
with the parameters $g_0$ and $g_1$. Actually, there is only one free parameter because we always want to
have a Minkowski vacuum in the potential. Therefore, we fix the parameter $g_0$ by demanding 
\begin{align}
  \partial _\varphi V(g_0)=\partial _\theta V(g_0)=V(g_0)=0~. \label{g0}
\end{align}

The corresponding potential $V$ with the linear $g$-function (\ref{ling}) reads
\begin{align}
\nonumber V=&\ \fracmm{M^2r^2}{3(1-r^2)^2}(1+r^2-2r\cos\theta)\\
&-\fracmm{M^2}{24(1-r^2)^2}(1+r^2-2r\cos\theta)^2\left[A+B{\rm{cos}}\theta+C{\rm{cos}}2\theta\right], \label{V_g_linear}
\end{align}
where we have introduced the notation
\begin{align}
A=6g_0g_1+g_1^2(-1-12r^2+r^4),\ \ B=-12g_0g_1r+4g_1^2r(2+r^2),\ \ C=6g_0g_1r^2,
\end{align}
and have used $r=\tanh\fracmm{\varphi}{\sqrt{6}}$. The first line of Eq.~$\eqref{V_g_linear}$ corresponds to the Starobinsky potential, whereas the second line shows the corrections caused by the $g$-function. 

The second derivative of the potential with respect to $\theta$ is given by
\begin{align} \lb{ddlgeq}
        \left.\fracmm{d^2V}{d\theta^2}\right|_{\theta=0}=\fracmm{M^2r}{6(1-r^2)^2}\[4r^2+3(1-r)^4\left(g_1^2-3g_0g_1+2r(g_0g_1-g_1^2)\right)\]~. 
\end{align}
In order to have an instability in the $\theta$-direction after the Starobinsky inflation, the second derivative in Eq.~$\eqref{ddlgeq}$ should become negative at some value of $r$. It is only possible when the signs of the parameters $g_0$ and $g_1$ are opposite.

\begin{figure}[h]
        \begin{minipage}[t]{0.45\hsize}
          \includegraphics[keepaspectratio, scale=0.8]{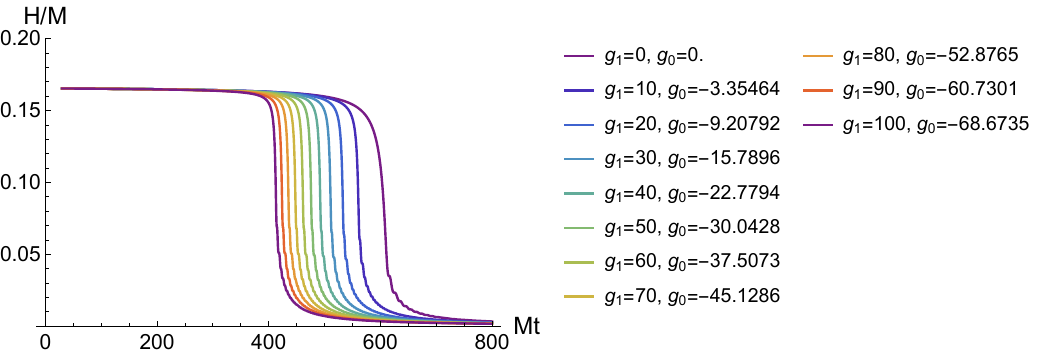}
        \end{minipage}
        \caption{The Hubble function for  selected values of $g_1$ and $g_0$. The initial conditions are $\theta(0)=0.001$ and $\varphi(0)=6$ with the vanishing initial velocities.}
        \label{lg_h}
\end{figure}

The potential $V$ in Eq.~$\eqref{V_g_linear}$ is symmetric with respect to interchange of signs of the parameters $g_0$ and $g_1$. We take $g_1$  positive and vary it from $0$ to $100$ with the increments of $10$ in order to get the shape of the Hubble function derived from Eqs.~(\ref{eq1}), (\ref{eq2}), (\ref{eq3}) and then determine whether double inflation occurs in each case. Having fixed the $g_1$, the $g_0$ is automatically fixed from Eq.~(\ref{g0}) by demanding the potential to be have a Minkowski vacuum. The results are shown in Fig.~\ref{lg_h} where the initial conditions are set as $\theta(0)=0.001$ and $\varphi(0)=6$ with the vanishing initial velocities.

The Hubble function in Fig.~2 has only one plateau,  which means no double inflation and no ultra-slow-roll phase in that case. Therefore, a more general $g$-function is needed.

\subsection{A linear $\tilde{g}$-function}

Since the functions $g$ and $\tilde{g}$ are non-linearly related, let us also consider the case of a linear $\tilde{g}$-function,  
\begin{align} \lb{tildeg}
        \tilde{g}(Z)=M(\tilde{g_0}+\tilde{g_1}Z),
\end{align}
with the coefficients $\tilde{g_0}$ and $\tilde{g_1}$. With $r= {\rm{tanh}}\fracmm{\varphi}{\sqrt{6}}$ and Eq.~(\ref{tildeg}), the potential $V$ is given by
\begin{align}
\nonumber V=\ &\fracmm{M^2r^2}{3(1-r^2)^2}(1+r^2-2r\cos\theta)\\
&-\fracmm{M^2}{3(1-r^2)^2}\left[9\tilde{g}_0^2+\tilde{g}_1^2(-1+4r^2)+12\tilde{g}_0\tilde{g}_1r {\rm{cos}}\theta\right].\lb{ltil_v}
\end{align}
We get a Minkowski vacuum by choosing 
$\tilde{g_0}$ properly.
The second derivative of the potential with respect to $\theta$ reads
\begin{align}
        \left.\fracmm{d^2V}{d\theta^2}\right|_{\theta=0}=\fracmm{2M^2r}{3(1-r^2)^2}(r^2+6\tilde{g_0}\tilde{g_1})
\end{align}
without using any approximation. Hence, the condition $-1/6<\tilde{g_0}\tilde{g_1}<0$ is necessary for having a $\theta$-instability.

\begin{figure}[h]
        \begin{minipage}[t]{0.45\hsize}
          \includegraphics[keepaspectratio, scale=0.8]{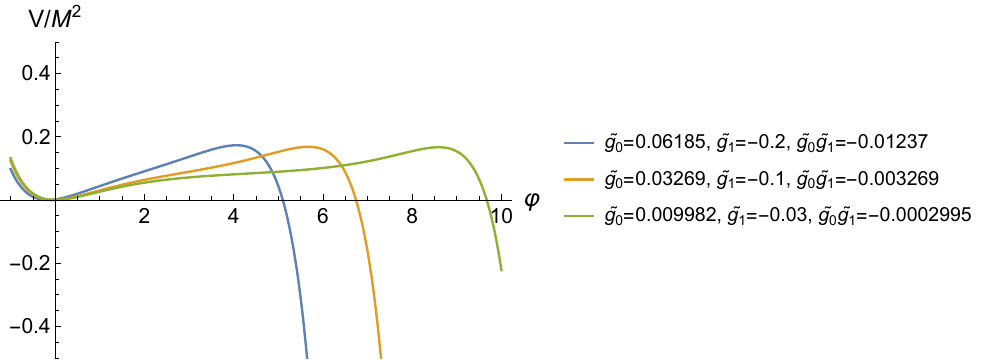}
        \end{minipage}
        \caption{The potential  at $\theta=0$  for $\tilde{g_1}$=-0.2, -0.1, -0.03.}
        \label{ltil_v_phi}
\end{figure}

\begin{figure}[h]
        \begin{minipage}[t]{0.45\hsize}
          \includegraphics[keepaspectratio, scale=0.8]{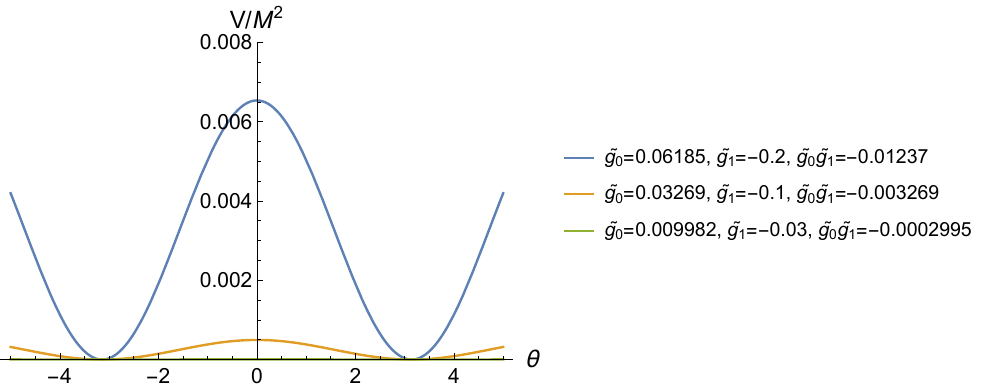}
        \end{minipage}
        \caption{The potential at $\varphi=\mathrm{const.}$  for $\tilde{g_1}$=-0.2, -0.1, -0.03.}
        \label{ltil_v_the}
\end{figure}

To give our examples, we choose three values of the parameter $\tilde{g_1}$ as $-0.2$, $-0.1$ and $-0.03$. The sign of $g_0$ should be positive in all these cases. The potential is the same after interchanging the signs between $\tilde{g_0}$ and $\tilde{g_1}$, which can be seen from Eq.~(\ref{ltil_v}). The shapes of the potential at $\theta=0$ and $\varphi=\mathrm{const.}$ near the minimum of the potential are shown in Figs.~\ref{ltil_v_phi} and \ref{ltil_v_the}, respectively.~\footnote{Since the Starobinsky slow roll inflation takes place for the $\varphi$-values between 
$5$ and $0.5$ \cite{Ketov:2021fww,Ivanov:2021chn}, we are not concerned by instabilities in the trans-Planckian region where we expect new physics  and our approach does not apply.}

As is clear from Fig.~\ref{ltil_v_phi}, when the value of the parameter $|\tilde{g_1}|$ gets larger, the length of the plateau of the potential gets shorter, so it becomes harder to have enough e-folds. On the other hand, Fig.~\ref{ltil_v_the} shows that the peak gets smaller when the  parameter $|\tilde{g_1}|$ gets smaller. Therefore, in order to have enough e-folds for the first stage of the inflation, the value of the parameter $|\tilde{g_1}|$ should be smaller or, in other words, closer to zero. However, when the parameter $\tilde{g_1}$ approaches zero, the $\theta$-instability will vanish, which results in the absence of the second stage of the inflation.

\begin{figure}[h]
        \begin{minipage}[t]{0.45\hsize}
          \includegraphics[keepaspectratio, scale=0.8]{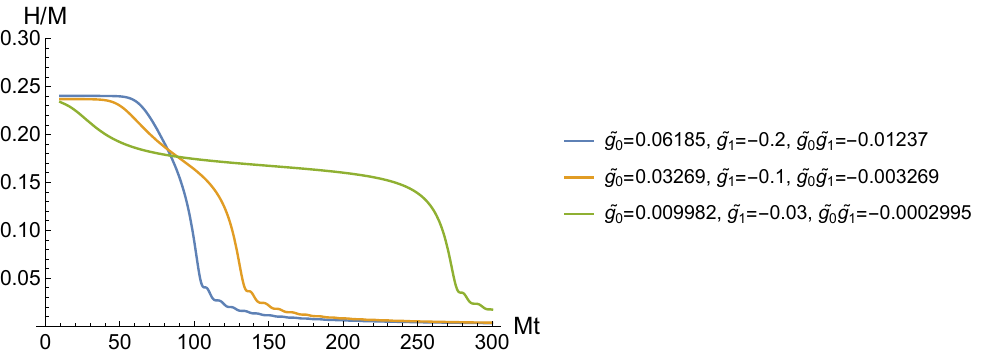}
        \end{minipage}
        \caption{The Hubble function in our examples with the initial conditions $\theta(0)=0.001$ and $\varphi(0)$, near the maximal value of the potential at $\theta=0$.}
        \label{ltil_h}
\end{figure}

The above arguments also can be checked directly by solving the equations of motion in the Friedman universe and inspecting the shape of the Hubble function. The shape of the Hubble function obtained from a solution to Eqs.~(\ref{eq1}), (\ref{eq2}) and (\ref{eq3}) is shown in Fig.~\ref{ltil_h}  where we have set the initial conditions as follows: the $\varphi(0)$ is the value at the maximum of the potential at $\theta=0$,~\footnote{Actually, the value of $\varphi(0)$ must be much smaller than the field value at the maximum, in order to give an impetus to the field (without it the field would stay at the maximum). The impetus slightly changes the shape of the Hubble function but does not change our conclusion.} with $\theta(0)=0.001$ and the vanishing initial velocities.

As is clear from Fig.~\ref{ltil_h}, the required duration of inflation can be achieved by lowering the value of $|\tilde{g_1}|$. However, having two plateaus cannot be achieved in all cases. When the parameters approach zero, the Hubble function apparently recovers the same shape as in the Starobinsky inflation. Therefore, we conclude that in the linear $\tilde{g}$-case double inflation cannot be achieved also, and thus a more general $\tilde{g}$-function is needed.

\section{The model with a quadratic $g$-function}

\begin{figure}[h]
  \begin{tabular}{ll}
    \begin{minipage}[t]{0.4\hsize}
      \centering
      \includegraphics[keepaspectratio, scale=0.7]{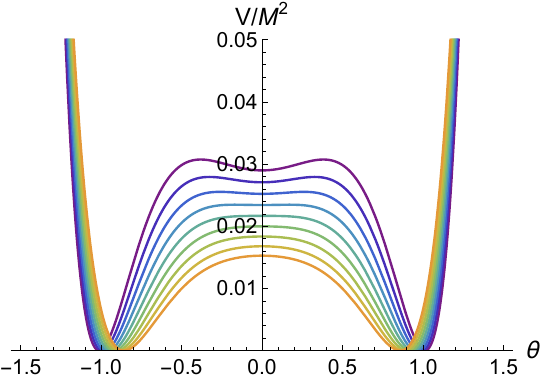}
      \subcaption{}
      \label{grg2}
    \end{minipage} &
    \begin{minipage}[t]{0.4\hsize}
      \centering
      \includegraphics[keepaspectratio, scale=0.7]{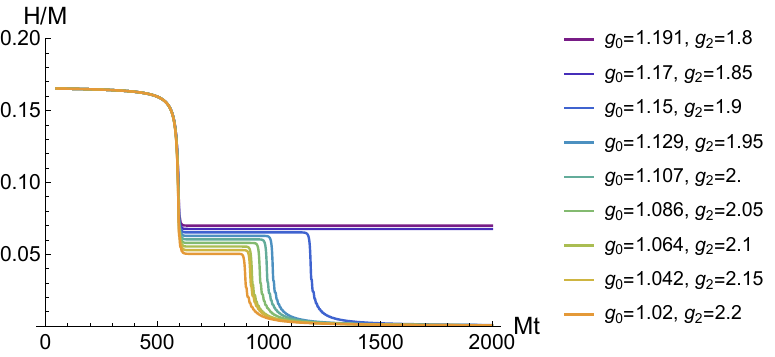}
      \subcaption{}
      \label{grg2h}
    \end{minipage}
  \end{tabular}
  \caption{(a) The sliced shape of the potential at fixed  $\varphi$ passing through the minima of the potential. (b) The Hubble function for $g_1=-1.0$ and the values of $g_2$ between $1.8$ and $2.2$. The initial values are set to $\varphi(0)=6$ and $ \theta(0)=0.001$ with the initial velocities all set to zero. The legend on the right shows the values of the parameters $g_0$ and $g_2$.}
  \label{grg2all}
\end{figure}

Motivated by the negative findings in the previous Section, in this Section we take a quadratic $g$ function,
\begin{equation} \label{quadg}
g(Z)=M(g_0+g_1Z+g_2Z^2)~,
\end{equation}
with three parameters $g_0$, $g_1$ and $g_2$, where the parameters $g_1$ and $g_2$ are arbitrary and $g_0$ is fixed by demanding Minkowski vacua, see  Fig.~\ref{grg2all}.

We search for the proper values of the parameters in order to obtain double inflation. By fixing one of the free parameters, the preferable shape of the potential can be obtained by varying another free parameter.  Increasing the value of the parameters causes the two minima of the potential to become closer to each other along the $\theta$ axis, which shortens duration of the second stage of inflation. 

The existing flexibility in the choice of the parameters is demonstrated by Fig.~\ref{grg2} that shows the slices of the potential at fixed 
$\varphi$, which cross the minima of the potential. In Fig.~\ref{grg2all} we have set $g_1=-1.0$ and have changed $g_2$ from $1.8$ to $2.2$ with the increment of $0.1$. The $g_0$ is derived by solving Eq.~(\ref{g0}). Though a slice does not show the shape of the potential at the inflection point, it does demonstrate that an instability can be achieved by increasing the value of the parameter $g_2$, while the maximum becomes lower and eventually becomes flat. Therefore, there should be the upper and lower bounds on $g_2$ for each possible $g_1$. We can check the actual range of the parameter $g_2$ by solving Eqs.~(\ref{eq1}), (\ref{eq2}) and (\ref{eq3}), which allows us to get the shape of the Hubble function. We set the initial values as $\varphi(0)=6$ and $\theta(0)=0.001$, with the vanishing initial velocities. Our results are shown in Fig.~\ref{grg2h}. The straight line after $Mt=600$, as in the $g_2=1.8$ and $g_2=1.85$ case, shows that an instability cannot be achieved and the trajectory did not fall to a minimum. As can be seen from Fig.~\ref{grg2h}, double inflation can be achieved for $1.9\leq g_2\leq2.2$.

\begin{figure}[h]
  \begin{tabular}{ll}
    \begin{minipage}[t]{0.4\hsize}
      \centering
      \includegraphics[keepaspectratio, scale=0.7]{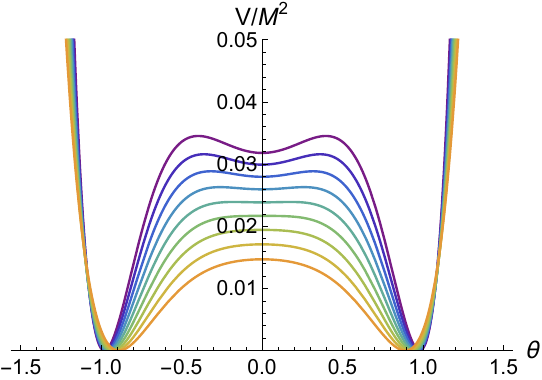}
      \subcaption{}
      \label{grg1}
    \end{minipage} &
    \begin{minipage}[t]{0.4\hsize}
      \centering
      \includegraphics[keepaspectratio, scale=0.7]{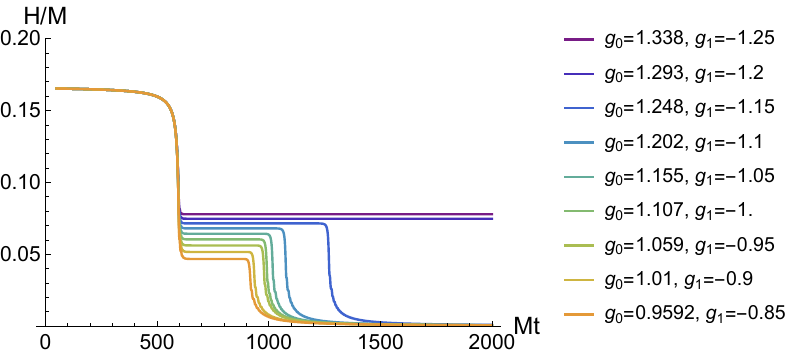}
      \subcaption{}
      \label{grg1h}
    \end{minipage}
  \end{tabular}
  \caption{(a) The sliced shape of the potential at fixed $\varphi$ passing through the minima of the potential. (b) The Hubble function for $g_2=2.0$ and the values of $g_1$ from $-1.25$ to $-0.85$. The initial values are set to $\varphi(0)=6$ and $\theta(0)=0.001$ with the vanishing initial velocities. The legends on the right shows the values of the parameters $g_0$ and $g_1$.}
  \label{grg1all}
\end{figure}

We also tested the case with fixed $g_2$ by changing $g_1$. Figure~\ref{grg1all} shows the case with $g_2=2.0$ and $g_1$ varied between $-1.25$ and $-0.85$, with all other conditions being the same as above. It reveals the same trend, namely, when increasing the value of $g_1$, a $\theta$-instability arises and the height of the potential gets smaller. In all cases, the value of $g_0$  decreases when $g_1$ and $g_2$ increase, while the potential gets closer to the one in the original Cecotti-Kallosh model of inflation \cite{Cecotti:2014ipa}.

\section{Inflationary trajectory, slow-roll parameters and \\ CMB observables}

Having derived Figs.~\ref{grg2all} and \ref{grg1all}, let us study the model with the fixed parameters $g_1=-1.0$, $g_2=2.0$ and $g_0=1.1073$, as a representative.

\begin{figure}[h]
  \begin{tabular}{cc}
    \begin{minipage}[t]{0.45\hsize}
      \centering
      \includegraphics[keepaspectratio, scale=0.8]{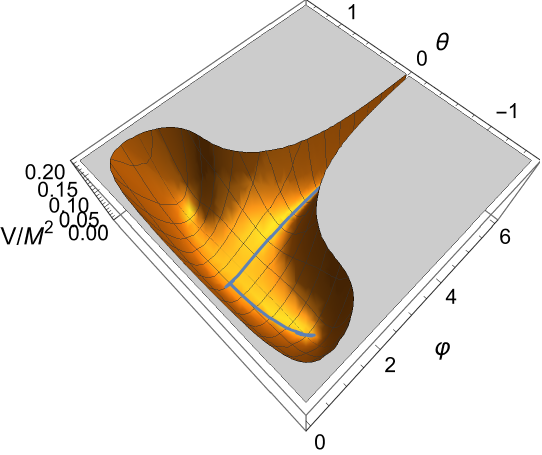}
      \subcaption{}
      \label{v2}
    \end{minipage} &
    \begin{minipage}[t]{0.45\hsize}
      \centering
      \includegraphics[keepaspectratio, scale=0.8]{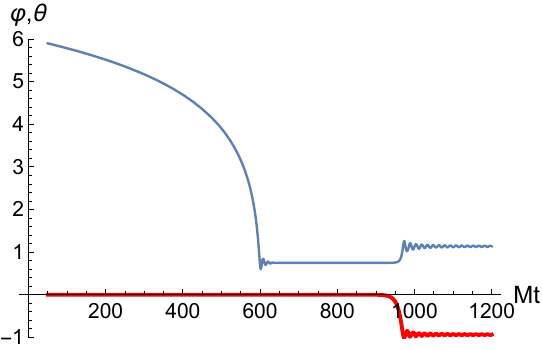}
      \subcaption{}
      \label{v2p}
    \end{minipage}
  \end{tabular}
  \caption{(a) The profile of the potential and the inflationary trajectory (blue). (b) The solutions $\varphi(t)$ (blue) and $\theta(t)$ (red) to the equations of motion. }
  \label{a3_1}
\end{figure}

\begin{figure}[h]
  \begin{tabular}{cc}
    \begin{minipage}[t]{0.45\hsize}
      \centering
      \includegraphics[keepaspectratio, scale=0.8]{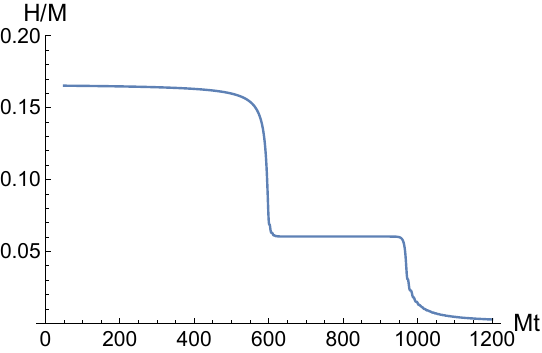}
      \subcaption{}
      \label{ah}
    \end{minipage} &
    \begin{minipage}[t]{0.45\hsize}
      \centering
      \includegraphics[keepaspectratio, scale=0.8]{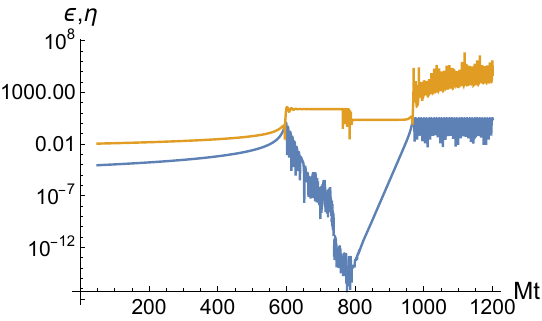}
      \subcaption{}
      \label{aslow}
    \end{minipage}
  \end{tabular}
  \caption{(a) The Hubble function. (b) The slow-roll parameters $\epsilon$ (blue) and $\eta$ (orange). }
  \label{a3_2}
\end{figure}

The scalar potential depending upon both fields $\varphi$ and $\theta$, the inflationary trajectory, and the time evolution of both fields in that potential are shown in Fig.~\ref{a3_1}. We also derived the Hubble function and the slow-roll parameters $\epsilon$ and $\eta$ in the representative model, which are shown in Fig.~\ref{a3_2}. The initial conditions are the same: $\varphi(0)=6$, $\theta(0)=1.0\times 10^{-4}$ and the vanishing initial velocities.

As is clear from Fig.~\ref{a3_1}, inflation begins as the single-field slow-roll Starobinsky inflation with $\varphi$ as the inflaton (the first stage of double
inflation). When the inflationary trajectory approaches the saddle point where one of the scalars gets a negative mass squared (the signal of tachyonic instability), the inflationary trajectory sharply changes its direction. After that inflation is driven by $\theta$-field as the inflaton (the second stage of double inflation). 

As is clear from Fig.~\ref{a3_2}, there is double inflation indeed because of two plateaus in the profile of the Hubble function, while the slow-roll
parameter $\epsilon$ becomes very small, which indicates the presence of the very short ultra-slow-roll phase between the two slow-roll stages.  The slow-roll conditions are violated in the ultra-slow-roll regime.

We define the start of the second stage of inflation (or the end of the first stage of inflation) by the time when the parameter $\eta$ first approaches one (or when the parameter $\epsilon$ first reaches its maximum), and the end of the second stage as the time when the parameter $\epsilon$ again reaches one first. The duration of the second stage of the inflation $\Delta N_2$ appears to be approximately 22.24 e-folds.

The main (CMB) inflationary observables are given by the spectral index $n_s$ and tensor-to-scalar ratio $r$ of scalar and tensor perturbations, respectively. We derived them in order to confront our model against Planck observational results. 

The slow-roll parameters $\epsilon$ and $\eta_{\Sigma\Sigma}$ are defined by ($\phi^A\equiv\{\varphi, \theta\}$)
\begin{align}
  &\epsilon\equiv-\fracmm{\dot{H}}{H^2}~,\quad \eta_{\Sigma\Sigma}\equiv\fracmm{\mathcal{M}_B^A\Sigma_A\Sigma^B}{V}~~~,\\
  &\Sigma^A\equiv\fracmm{\dot{\phi}^A}{|\dot{\phi}|}~,\quad \mathcal{M}_B^A\equiv G^{AC}\nabla_B\partial_CV-\mathrm{R}^A_{~CDB}\dot{\phi}^C\dot{\phi}^D~,
\end{align}
in terms of the metric $G_{AC}$ in the scalar field space and its Riemann-Christoffel curvature $\mathrm{R}^A_{~CDB}$. The isocuvature parameter is defined by 
\begin{align}
  \eta_{\Omega\Omega}\equiv\fracmm{\mathcal{M}_B^A\Omega_A\Omega^B}{V}~, \quad
  \mathrm{where} \quad \Omega^A\equiv\fracmm{{\omega}^A}{|{\omega}|}~, \quad
  \omega\equiv\dot{\Sigma^A}+\Gamma^A_{BC}\Sigma^B\dot{\phi}^C~.
\end{align}
The transfer function are defined by
\begin{align}
  T_{SS}(t_1,t_2)\equiv \exp \left[\int_{t_2}^{t_1}dt'\beta(t')H(t')\right]~,\\
  T_{RS}(t_1,t_2)\equiv 2\int_{t_2}^{t_1}dt'|\omega(t')|T_{SS}(t_1,t_2)~,\\
  \mathrm{where} \quad \beta(t)\equiv-2\epsilon+\eta_{\Sigma\Sigma}-\eta_{\Omega\Omega}-\fracmm{4|\omega|^2}{3H^2}~.
\end{align}

These equations are enough to compute the observables by applying the Mathematica packages for numerical calculations with the transport method in multi-field models of inflation \cite{Dias:2015rca}. In particular,  the observables $n_s$ and $r$ are given by
\begin{align}
  n_s=1-6\epsilon+2\eta_{\Sigma\Sigma} \quad {\rm and} \quad r=\fracmm{16\epsilon}{1+T_{RS}^2}~~.
\end{align}

The values of the observables in our representative model at the CMB pivot scale, which corresponds to $k_{\rm CMB}=0.05~{\rm Mpc}^{-1}$ (we set 70 e-folds as the duration of the whole double inflation), and the same initial values mentioned above are shown in Table~\ref{a2_tab2} together with the values of $g_0$, the values of $\varphi$ and $\theta$ at the minimum, and the number of e-folds for the second stage of inflation. Table~\ref{a2_tab2} also contains the values of the other parameters,  $g_1$ and $g_2$.

The improved precision measurements of the CMB radiation \cite{BICEP:2021xfz,Tristram:2021tvh} give for the spectral tilt $n_s$ of scalar perturbations
the values 
\be \lb{ns}
n_s = 0.9649 \pm 0.0042 \quad (68\%~{\rm C.L.})
\ee
The current observational upper bound \cite{BICEP:2021xfz,Tristram:2021tvh} on the CMB tensor-to-scalar ratio $r$ is given by
\be \lb{rplanck}
r < 0.036\quad (95\%~{\rm C.L.})
\ee

As is clear from Table~\ref{a2_tab2}, the values of $r$ in our model are well below the observational bound but the values of $n_s$ are outside
the $1\sigma$ range for the given values of $\Delta N_2$, except the case of $g_1=-0.25$. The reason for that is our intension to maximize the value of $\Delta N_2$ which is crucial 
for larger PBH masses, see Eq.~(\ref{pbhm}) below. The $1\sigma$ agreement can be achieved by changing the parameters at the expense of decreasing $\Delta N_2$ to 10.~\footnote{For instance, it can be easily achieved by changing the value of $g_2$.} Nevertheless, it also follows from Table~\ref{a2_tab2} that the $3\sigma$ agreement with the observed value of $n_s$ is still possible for $\Delta N_2\leq 22.24$.

\begin{table}[ht]
  \centering
  \begin{tabular}{l r r r r r}
  \toprule
  $g_1$ & $-2.0$ & $-1.5$ & $-1.0$ & $-0.5$ & $-0.25$ \\
  $g_2$ & $2.9$ & $2.5$ & $2.0$ & $1.4$  & $1.05$ \\
  \midrule
  $g_0$           & $1.6292$ & $1.35754$ & $1.1073$ & $0.862908$ & $0.728364$\\
  $\varphi_{min}$ & $1.33857$ & $1.23483$ & $1.13543$ & $0.999405$  & $0.752367$\\
  $\theta_{min}$ & $0.832134$ & $0.869356$ & $0.938336$ & $1.05348$ & $1.03411$ \\
  $\Delta N_2$     & $31.50$ & $25.99$ & $22.24$ & $20.31$ & $12.69$ \\
  $n_s$         & $0.9416$ & $0.9496$ & $0.9537$ & $0.9551$ & $0.9616$ \\
  $r$          & $0.0092$ & $0.0069$ & $0.0059$ & $0.0055$ & $0.0041$\\
    \bottomrule
  \end{tabular}
  \captionsetup{width=.9\linewidth}
  \caption{The values of the parameters $(g_1, g_2, g_0)$, the field values ($\varphi_{min}, \theta_{min}$) at the minimum, the number of e-foldings for the second stage of the inflation ($\Delta N_2$), and the CMB observables $(n_s, r)$. The initial values are $\varphi(0)=6$ and $\theta(0)=10^{-4}$, with the vanishing initial velocities.}
\label{a2_tab2}
\end{table}

We found difficult to derive the power spectrum in our models due to field oscillations along $\theta=0$ because our numerical calculations
failed. This "primordial feature" is likely caused due to resonances during inflation on sub-horizon scales. Such resonances can occur when
a massive scalar field oscillates at the bottom of the potential \cite{Pahud:2008ae,Chluba:2015bqa,Fumagalli:2021cel}, like in our model. To avoid this problem,
we computed the power spectrum in our model by considering the first and second stages of double inflation separately, see the next Sections.
However, it required setting initial conditions at the beginning of the second stage of inflation, which we did not determine and, hence, took them
randomly.

\section{Dependence upon initial conditions} 

In this Section, we take a closer look on dependence of the observables upon initial conditions.  There are two independent sets of initial conditions for sinflaton $\theta$ and inflaton $\varphi$. In the Starobinsky-like inflation, the power spectrum is more sensitive to initial conditions on $\theta$ rather than those on $\varphi$ because we deal with the  attractor model of inflation where the dependence upon the inflaton initial conditions is suppressed (besides duration of inflation). Our numerical calculations confirm this expectation. That is why we focus on dependence upon initial conditions on $\theta$ and field dynamics near the critical point.

As regards the primordial power spectrum, see the next Section, it is not affected by initial conditions on adiabatic and isocurvature perturbations. It happens because the power spectrum enhancement is determined by behavior of field perturbations near the critical point that is independent upon initial conditions. The evolution of perturbations, obtained by numerical calculations in our model, appears to be usual (as expected) after integrating the equations governing adiabatic and isocurvature perturbations with the isocurvature pumping mechanism, as was described e.g., in Ref.~\cite{Gundhi:2020kzm}, Sections VI and VII.

\begin{figure}[h]
  \centering
  \begin{minipage}[t]{0.45\hsize}
    \includegraphics[keepaspectratio, scale=0.8]{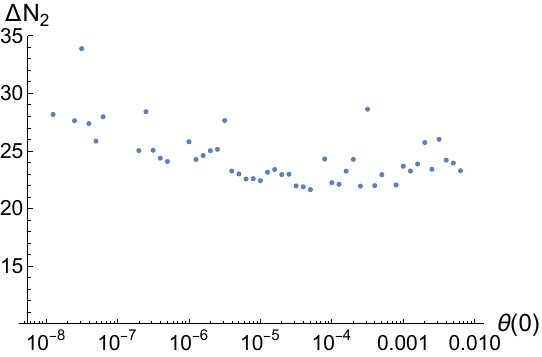}
  \end{minipage}
  \caption{The dependence of the duration $\Delta N_2$ of the second stage of inflation upon the value of initial condition $\theta(0)$ in our
  representative model.}
  \label{a2_vel}
\end{figure}

\begin{figure}[h]
  \begin{tabular}{ccc}
    \begin{minipage}[t]{0.45\hsize}
      \centering
      \includegraphics[keepaspectratio, scale=0.8]{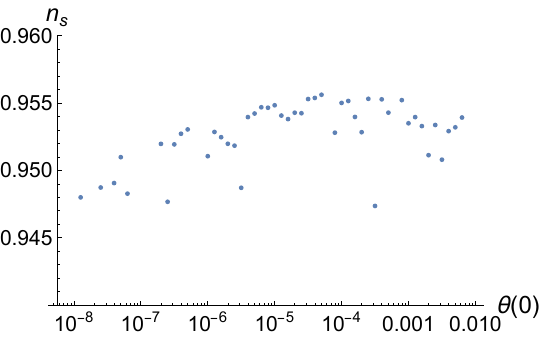}
      \subcaption{}
      \label{ah}
    \end{minipage} &
    \begin{minipage}[t]{0.45\hsize}
      \centering
      \includegraphics[keepaspectratio, scale=0.8]{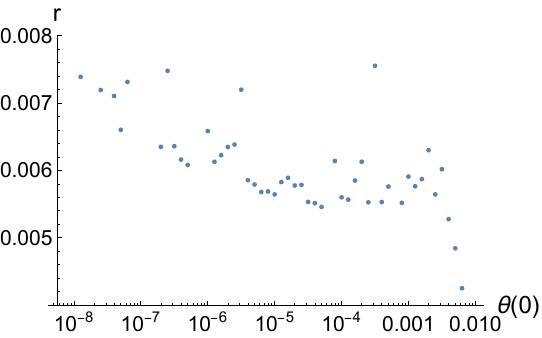}
      \subcaption{}
      \label{aslow}
    \end{minipage} 
  \end{tabular}
  \caption{The dependence of $n_s$ and $r$ upon the initial condition $\theta(0)$ in our representative model.}
  \label{a2_ini_ob}
\end{figure}

In order to get the dependence upon the initial value of $\theta$, we numerically computed the duration ($\Delta N_2$) of the second stage of inflation (in e-folds) in the representative model for the different values of $\theta(0)$ between $10^{-8}$ and $10^{-2}$. Our numerical results are shown in Fig.~\ref{a2_vel}. As is clear from Fig.~\ref{a2_vel}, the values of $\Delta N_2$ weakly depend upon small changes in the initial value of $\theta$.  The values of $\Delta N_2$ vary from about $20$ to $30$ that changes the corresponding values of the observables by the order of $10^{-2}$.  The similar results for the CMB observables are given  in Fig.~\ref{a2_ini_ob}, which are a bit more sensitive to initial conditions.

The $\theta$-field is strongly suppressed at its minimum during the first stage of the inflation but becomes sensitive to any additional small impact at the saddle point where quantum corrections (for example, due to quantum diffusion \cite{Pattison:2021oen}) may become important while the impact of initial conditions becomes negligible. Therefore, we study the first and second stages of inflation separately, and assume a small non-vanishing value (kick) of 
the $\theta$-velocity at the beginning of the second stage of inflation. We denote  $t_1$ to be the time at the end of the first stage inflation, and 
 $t_2$ to be the time of the end of the second stage of inflation. As regards the first stage of inflation, we use our representative model studied in 
 the preceding Sections with the same initial value for $\varphi$ in order to get the resulting values of $\varphi(t)$ and $\theta(t)$ and their velocities at the end of the first stage of inflation.

\begin{figure}[h]
  \centering
  \begin{minipage}[t]{0.45\hsize}
    \includegraphics[keepaspectratio, scale=0.8]{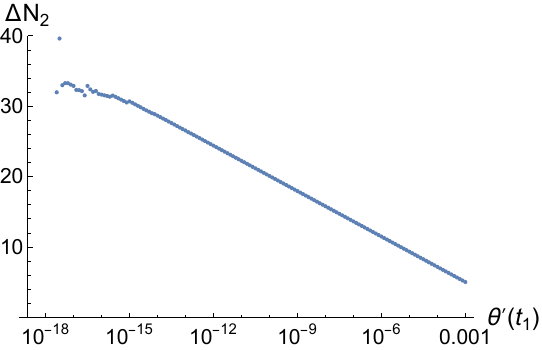}
  \end{minipage}
  \caption{The dependence of duration of the second stage of inflation, $\Delta N_2$, upon of the initial condition $\theta'(t_1)$ with 
  the parameters $(g_0, g_1, g_2)=(1.1496, -1.0, 1.9)$.}
  \label{a2_ini}
\end{figure}

Next, we check whether we can get enough e-folds $\Delta N_2$ by adding a small $\theta$-velocity at the saddle point. In our representative model we use the parameters $(g_0, g_1 ,g_2)=(1.1496, -1.0, 1.9)$ and the initial values $(\varphi(0), \theta(0))=(6, 0)$ with the vanishing initial velocities. Then the values of $\varphi(t)$ and $\varphi'(t)$ at the saddle point are $|\varphi(t)|\sim 0.86$ and $\varphi'(t)\lesssim\mathcal{O}(0.1)$, respectively. Our result for the dependence of $\Delta N_2$ upon the kick velocity $\theta'(t_1)$ at the saddle point is shown in Fig.~\ref{a2_ini}. As is clear from 
Fig.~\ref{a2_ini},  $\Delta N_2\sim20$ can be achieved with $\theta'(t_1)\sim\mathcal{O}(10^{-10})$. 

Having obtained those results, we derived the corresponding CMB observables (with the total duration of inflation given by 70 e-folds) as follows:
\begin{align}
  n_s=0.9634~, \quad r=0.0038~.
\end{align}

\section{Power spectrum and PBH masses}

In this Section we numerically compute the power spectrum of scalar (curvature) perturbations by using the transport method with the Mathematica package \cite{Dias:2015rca} that was widely used in the literature. In order to efficiently produce the primordial black holes, the enhancement (peak) in the power spectrum against its CMB value should be $10^6$ times at least. We do not provide here the underlying equations, defining the power spectrum and the  "isocurvature pumping" amplification mechanism, because they are available in the literature, see e.g., subsection
III.B of  Ref.~\cite{Gundhi:2020kzm} for details.

Given a peak, the masses of generated PBHs can be calculated by the equation \cite{Pi:2017gih}
\begin{align} \label{pbhm}
  M_{\mathrm{PBH}}\simeq \fracmm{M_{\mathrm{Pl}}^2}{H(t_*)}\mathrm{exp}\left[2(N_{\mathrm{end}}-N_*)+\int_{t_*}^{t_{\mathrm{exit}}}\epsilon(t)H(t)dt\right]~,
\end{align}
where $t_*$ and $N_*$ are the time and e-folds at the end of the first stage of the inflation, $N_{\mathrm{end}}$ is 
the number of e-folds at the end of (the second stage) of inflation, and $t_{\mathrm{exit}}$ is the time when the CMB pivot scale exits the horizon. Equation (\ref{pbhm})  was derived in Ref.~\cite{Pi:2017gih} by estimating the PBH mass inside the horizon, when the mode corresponding to the peak in the power spectrum re-entered the horizon. As is clear from Eq.~(\ref{pbhm}), the PBH masses are very sensitive to the value of $\D N_2=(N_{\mathrm{end}}-N_* )$.

\begin{figure}[h]
  \centering
  \begin{minipage}[t]{0.45\hsize}
    \includegraphics[keepaspectratio, scale=0.8]{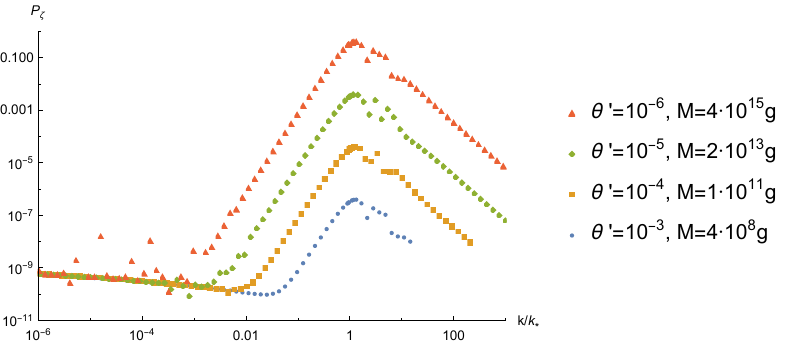}
  \end{minipage}
  \caption{The power spectrum of scalar perturbations in the case of $g_1=1.0, g_2=1.9$ for the different values of $\theta'(t_1)$. The initial conditions are $\varphi(0)=7, \theta(0)=0$ with the vanishing initial velocities. The pivot scale $k_{*}$ corresponds to the end of the first stage of inflation. The top spectrum has significant oscillations, indicating a breakdown of our numerical calculations, so  we discarded it in favor of the others. }
  \label{a2_pow1}
\end{figure}

Our first example is the case with $g_1=1.0, g_2=1.9$, where we have scanned the values of the $\theta$-velocity at the saddle point  from $10^{-3}$ to $10^{-6}$. The power spectrum is shown in Fig.~\ref{a2_pow1}. As is clear from  Fig.~\ref{a2_pow1}, the enhancement of the power spectrum strongly depends on the value of $\theta'(t_1)$. In order to get the desired height of the peak from the CMB value of $10^{-9}$ to $10^{-2}$, we have to choose the value of $\theta'(t_1)$ within a small margin. We also give the corresponding PBH masses (in grams) in Fig.~\ref{a2_pow1}.

\begin{figure}[h]
  \centering
  \begin{minipage}[t]{0.45\hsize}
    \includegraphics[keepaspectratio, scale=0.8]{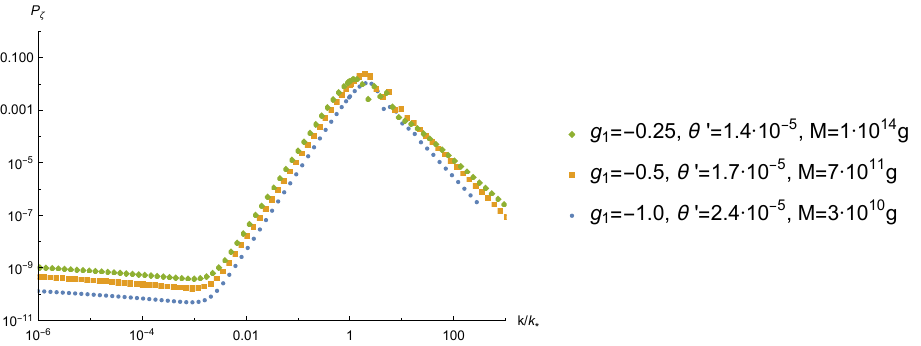}
  \end{minipage}
  \caption{The power spectrum of scalar perturbations for the different values of $g_1$ with the peak height at $0.01$. The initial conditions are   $\varphi(0)=7, \theta(0)=0$ with the vanishing initial velocities. The pivot scale $k_{*}$ corresponds to the end of the first stage of inflation.}
  \label{a2_pow2}
\end{figure}

As another example, we choose $g_1={-1.0, -0.5, -0.25}$ and adjust the proper values for $g_2$ and $g_0$. The $\theta'(t_1)$ is chosen to get
a peak up to $10^{-2}$ in the power spectrum. The resulting power spectrum is given in Fig.~\ref{a2_pow2}.

The parameter values together with the values of the related observables corresponding to Fig.~\ref{a2_pow2} is collected  in Table \ref{a2_tab3}.
As is clear from Table \ref{a2_tab3}, smaller values of $\abs{g_1}$ lead to larger PBHs masses.

\begin{table}[ht]
  \centering
  \begin{tabular}{c c c c c c c c c}
  \toprule
  $g_1$ & $g_0$ & $g_2$ & $\theta'$ & $\Delta N_2$ & $n_s$ & $r$ & $P_{enh}$ & $M_{\mathrm{PBH}}~(g)$ \\
  \midrule
  $-0.25$ & $0.7767$ & $0.98$ & $1.4\times 10^{-5}$ & $14.70$ & $0.9604$ & $0.0044$ & $1.9\times 10^{6}$ & $1.3\times 10^{14}$  \\
  $-0.5$ & $0.9106$ & $1.3$ & $1.7\times 10^{-5}$ & $12.32$ & $0.9623$ & $0.0040$ & $5.9\times 10^6$ & $7.4\times 10^{11}$  \\
  $-1.0$ & $1.1496$ & $1.9$ & $2.4\times 10^{-5}$ & $10.86$ & $0.9634$ & $0.0038$ & $1.1\times 10^7$ & $2.9\times 10^{10}$ \\
  \bottomrule
  \end{tabular}
  \captionsetup{width=.9\linewidth}
  \caption{The parameters, the CMB observables, the enhancement of the power spectrum, and the PBH masses in our models. The kick velocity  $\theta'(t_1)$ is chosen to have the peak summit at $10^{-2}$ in the power spectrum. The initial conditions are  $\varphi(0)=7$ and $\theta(0)=0$  with the vanishing initial velocities. The pivot scale is $k_{\rm CMB}=0.05~{\rm Mpc}^{-1}$. The total duration of inflation is 70 e-folds.}
  \label{a2_tab3}
\end{table}

To the end of this Section, we comment on the quadratic Ansatz for the $\tilde{g}$-function in Eq.~(\ref{pot}) as 
\begin{equation}\label{tildegquad}
\tilde{g}(Z)=M(\tilde{g}_0+\tilde{g}_1Z+\tilde{g}_2Z^2)
\end{equation}
in our search for proper values of its three parameters for double inflation.

We did not find the desired shape of the potential that would support double inflation in this case. To examine the potential, we have set a space with the three parameters $\tilde{g}_0$, $\tilde{g}_1$ and $\tilde{g}_2$ within the range from $-100$ to $+100$, and have imposed the following conditions on the potential: (i) the value of the potential must be positive, (ii) the duration of inflation should be long enough, (iii) the potential along the $\theta$-direction should be stable during the first stage of inflation, and (iv)  the height of the potential should be limited from above. By demanding these conditions on the parameter space, no area was found to satisfy them. Therefore, the model with the $\tilde{g}$-function
(\ref{tildegquad}) is ruled out. It is worth noticing that the four conditions above are rather mild, e.g., in the case of the quadratic $g$-function they are easily satisfied.

\section{Conclusion}

It follows from Table \ref{a2_tab3} that perfect consistency of our models with CMB measurements restricts the possible PBH masses by $10^{14}$ g
even after fine-tuning of the parameters, which is below the Hawking (quantum black hole) evaporation limit of  $10^{15}$ g. It means that those
PBH cannot be part of the current dark matter.

This in contrast to the different extensions of Starobinsky model in supergravity proposed and investigated in 
Refs.~\cite{Aldabergenov:2020bpt,Aldabergenov:2022rfc,Aldabergenov:2020yok} where the generated PBH masses can reach $10^{21}$ g.
It is also different from the non-supersymmetric Appleby-Battye-Starobinsky model \cite{Appleby:2009uf} adapted in Ref.~\cite{Frolovsky:2022ewg} for PBH production, where the PBH masses can reach $10^{19}$ g.

Of course, all the above goes under the assumption that the PBH do not get extra mass over time during the evolution of the Universe. 
It might be possible that inflation generates a lot of PBH with masses of $10^{14}$ g, which keep their masses and may get even larger masses via accretion and mergers. However, then one would need a lot of those PBH in order to account for a significant part of the current dark matter, which would contradict to the non-detection of extra-galactic gamma-rays from their Hawking radiation. 

We conclude that our supergravity models studied in this paper are consistent with CMB measurements and do lead to the efficient PBH production with the fine-tuned quadratic $g$-function in the superpotential, but those PBH cannot be part of  the current dark matter. We did not study more complicated $g$- and $\tilde{g}$-functions beyond quadratic polynomials and exponentials, because it could only be done on the case-by-case basis and would require a separate investigation. In the case of quadratic $g$-functions, which was  in the focus of this investigation, all the parameters had to be fixed in order to achieve the desired enhancement of the scalar power spectrum needed for PBH production. The obvious next step would be adding one extra parameter by studying cubic $g(Z)$-functions (our work in progress).

\section*{Acknowledgements}

SA was supported by Chung-Ang University and the Basic Science Research Program through the National Research Foundation (NRF) funded by the Ministry of Education, Science and Technology in South Korea under the grant 
No.~NRF-2022R1A2C2003567.  RI and SVK were supported by Tokyo Metropolitan University. SVK was also supported by the World Premier International Research Center Initiative (WPI Initiative), MEXT, Japan, the Japanese Society for Promotion of Science under the grant No.~22K03624, and the Tomsk Polytechnic University Development Program 
Priority-2030-NIP/EB-004-0000-2022. 

The authors are grateful to David I. Kaiser, Edward W. Kolb, Burt Ovrut, Misao Sasaki and two anonymous referees for discussions and correspondence.

\begin{appendix}

\section{Exponential $\tilde{g}$-function}

In Appendices A and B we provide more support for our choice of {\it polynomial} functions, $g(Z)$ and $\tilde{g}(Z)$, adopted  in the main body of this paper, by considering two alternatives (exponential- or KKLT-type \cite{Kachru:2003aw}) and
demonstrating that those alternatives should be ruled out because inflation is destabilized (too low e-folds).

\begin{figure}[t]
 \begin{minipage}{0.5\hsize}
  \begin{center}
   \includegraphics[width=70mm]{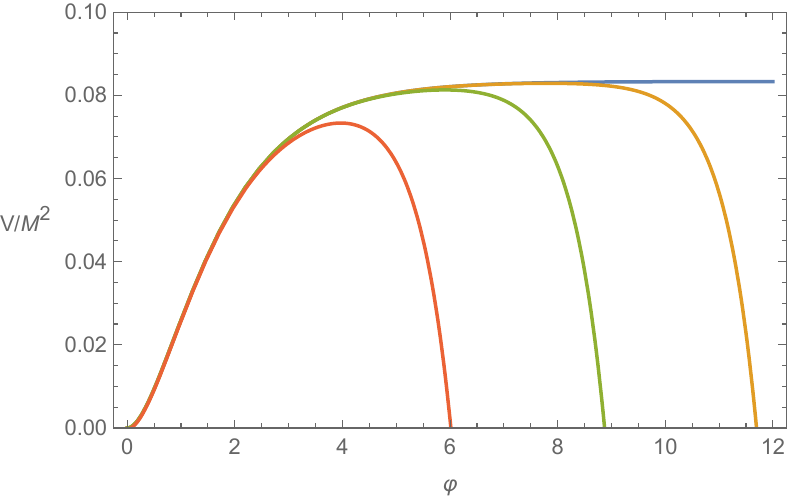}
  \end{center}
 \end{minipage}
 \begin{minipage}{0.5\hsize}
  \begin{center}
   \includegraphics[width=70mm]{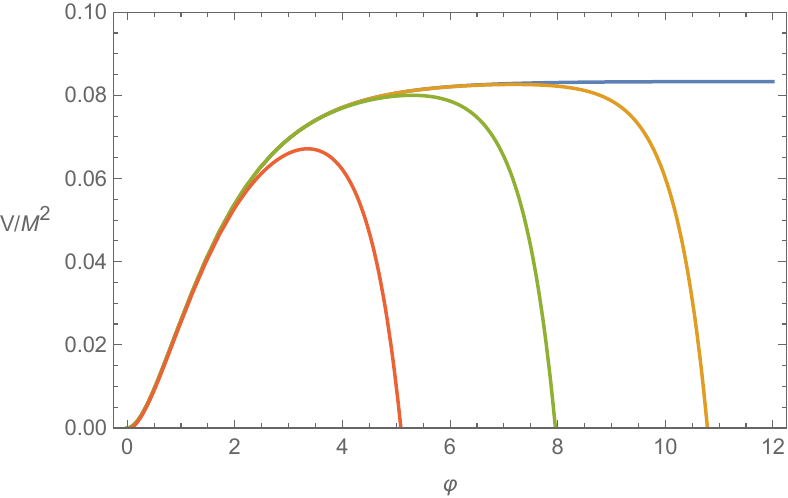}
  \end{center}
 \end{minipage}
\caption{Left: the effective scalar potential at $\theta=0$ in the model (\ref{gtil_e}).  The parameter $\tilde{A}$ is chosen to be $0$ (blue), $10^{-4}$ (yellow), $10^{-3}$ (green) and $10^{-2}$ (red), with $\tilde{C}=0$ and $\tilde{a}=1$ being fixed. Right: the parameter $\tilde{C}$ is chosen to be $0$ (blue), $10^{-4}$ (yellow), $10^{-3}$ (green) and $10^{-2}$ (red), with $\tilde{A}=0$.}
  \label{fig1}
\end{figure}

Let us choose
\begin{align}
\tilde{g}(Z)=M\left(\tilde{C}+\tilde{A} e^{-\tilde{a}Z}\right)~,\label{gtil_e}    
\end{align}
with the real parameters  $\tilde{C}, \tilde{A}$ and $\tilde{a}$. In Fig.~\ref{fig1}, we show the impact of 
$\tilde{C}$ and $\tilde{A}$ separately, when $\tilde{a}=1$ is fixed. As is clear from Fig.~\ref{fig1}, 
large values of $\tilde{C}$ and $\tilde{A}$ spoil the inflation trajectory at $\theta=0$.  The approximate upper bounds  on those parameters are very low,
\begin{align}
|\tilde{A}|, ~~|\tilde{C}|~~\lesssim 10^{-4}~.\label{UB}
\end{align}
These constraints become even more restrictive for negative values of $\tilde{a}$, while they also persist
when both coefficients $\tilde{A}$ and $\tilde{C}$ do not vanish. Within the 
range (\ref{UB}), the second derivative  $d^2V/d\theta^2$ always takes positive values, and there is no tachyonic instability in the $\theta$-direction. 

\section{Exponential $g$-function }

\begin{figure}[t]
 \begin{minipage}{0.5\hsize}
  \begin{center}
   \includegraphics[width=70mm]{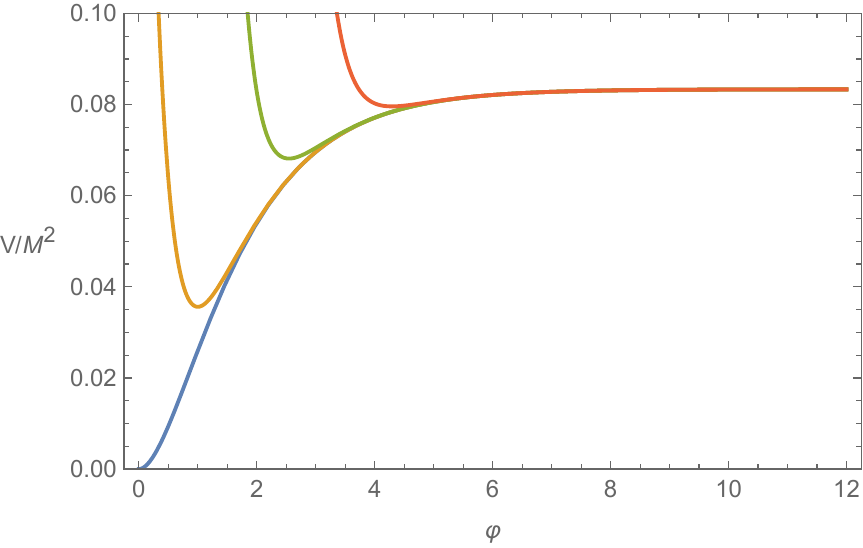}
  \end{center}
 \end{minipage}
 \begin{minipage}{0.5\hsize}
  \begin{center}
   \includegraphics[width=70mm]{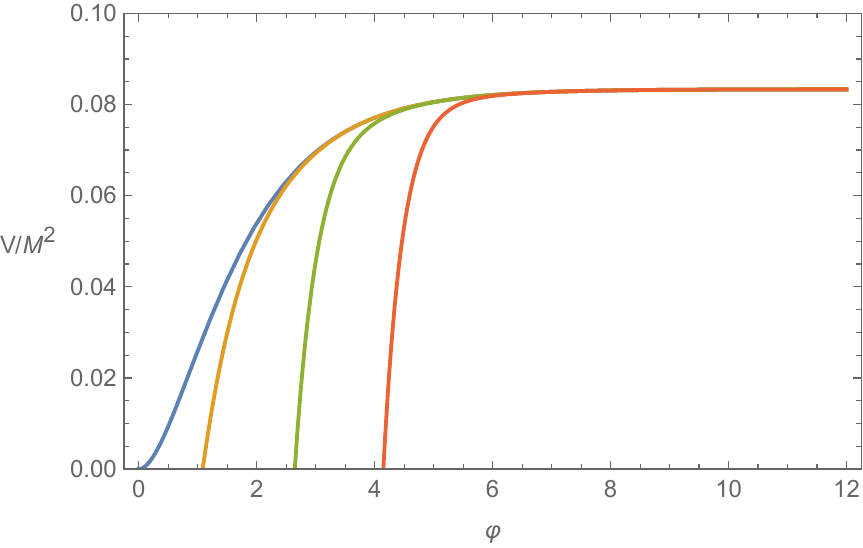}
  \end{center}
 \end{minipage}
 
  \begin{minipage}{0.5\hsize}
  \begin{center}
   \includegraphics[width=70mm]{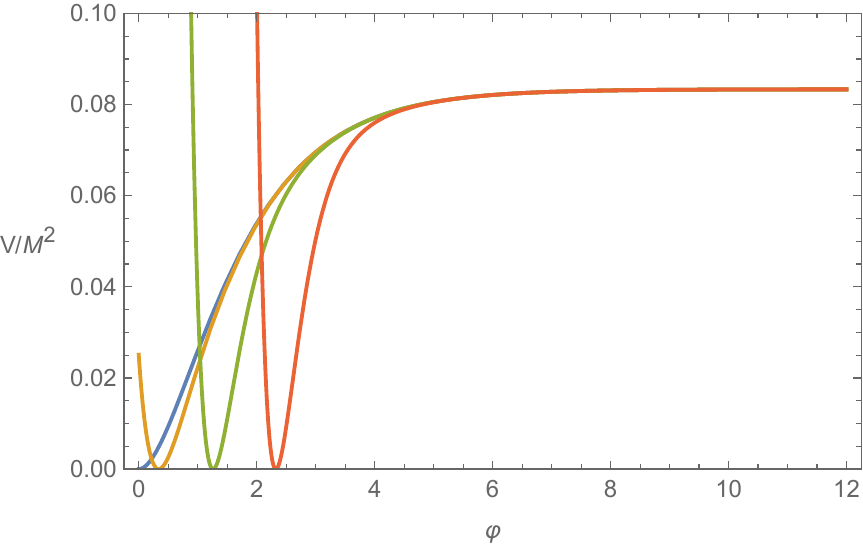}
  \end{center}
 \end{minipage}
 \begin{minipage}{0.5\hsize}
  \begin{center}
   \includegraphics[width=70mm]{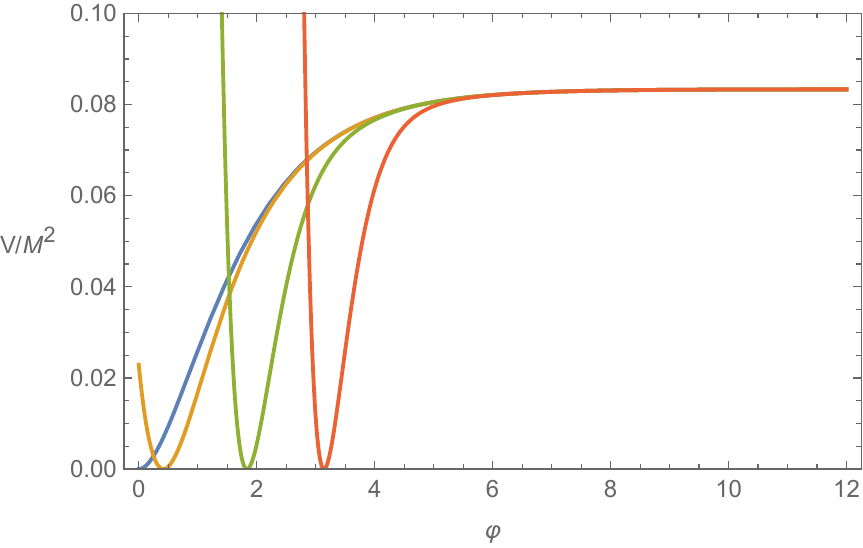}
  \end{center}
 \end{minipage}
 
  \begin{minipage}{0.5\hsize}
  \begin{center}
   \includegraphics[width=70mm]{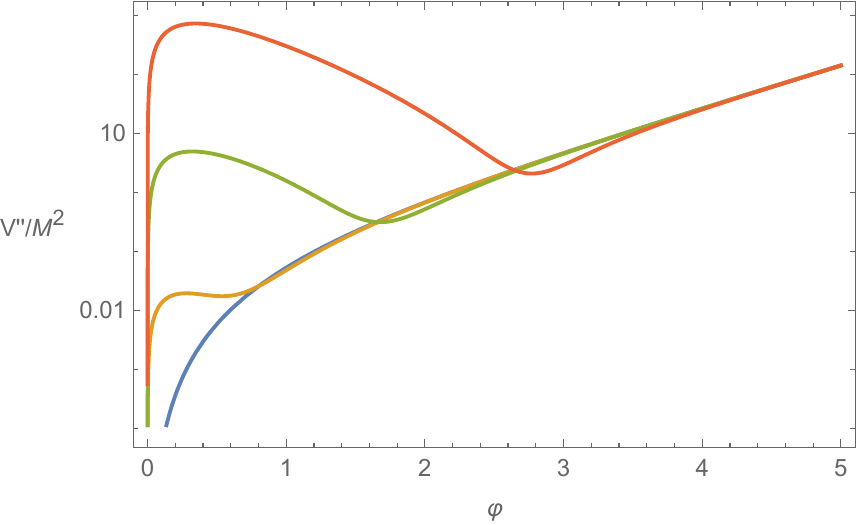}
  \end{center}
 \end{minipage}
 \begin{minipage}{0.5\hsize}
  \begin{center}
   \includegraphics[width=70mm]{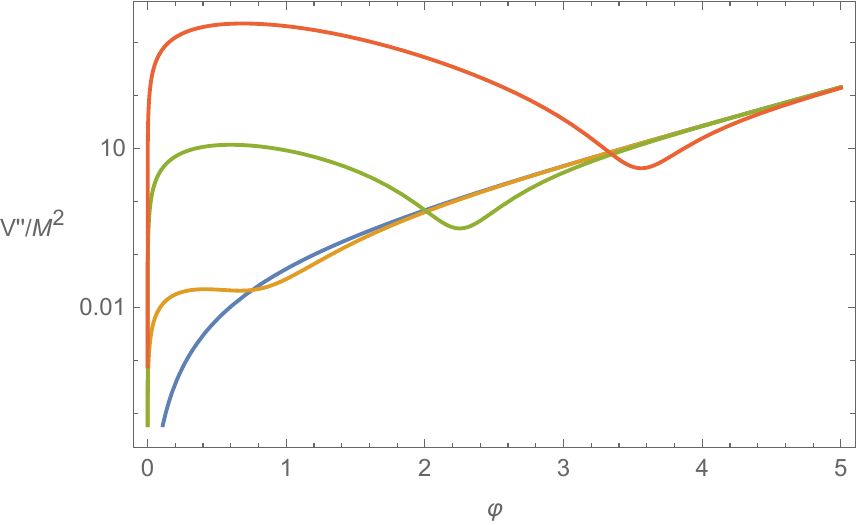}
  \end{center}
 \end{minipage}
\caption{Top left: the effective scalar potential at $\theta=0$ with $a=1$ and $C=0$. The parameter $A$ is chosen as $0$ (blue), $1$ (yellow), $10$ (green) and $100$ (red). Top right: the same as in the left figure but with $a=-1$. Middle left: the effective scalar potential at $\theta=0$, with $a=1$ and $C$ being tuned to realize a  Minkowski vacuum. The parameters are chosen as $(A,C)=(1,-1)$, $(A,C)=(10,-7.6)$ and $(A,C)=(100,-54)$. Middle right: the same as in the left figure but with $a=-1$. The parameters are chosen as $(A,C)=(1,-0.9)$, $(A,C)=(10,-15)$ and $(A,C)=(100,-210)$. Bottom left: the values of the second derivative  $d^2V/d\theta^2$ at $\theta=0$ with $a=1$ and the same parameter set as in the 
middle-left figure. Bottom right: the same as in the left figure but with $a=-1$.}
  \label{fig3}
\end{figure}

Here we consider an exponential-type $g$-function instead,
\begin{align}
g(Z)=M\left(C+A e^{-aZ}\right),\label{g_e}    
\end{align}
where all parameters are real. First, in the case $C=0$, we derive Fig.~\ref{fig3}. On the top of Fig.~\ref{fig3}, the inflation trajectory for small $\varphi$ around the minimum is destroyed by the $A$-term in
 Eq.~(\ref{g_e}), depending upon the sign of $a$. However, this can be cured when we include the parameter $C$ and tune it accordingly. This is contrast to the case in Appendix A with the function 
 $\tilde{g}$  defined by Eq.~(\ref{gtil_e}), where both parameters $\tilde{A}$ and $\tilde{C}$ affect the potential at large field values of $\varphi$, and the destabilization of the potential cannot be cured by adjusting 
 $\tilde{C}$.

In the middle of Fig.~\ref{fig3} we show the results of turning on the coefficient $C$. As one  can see in Fig.~\ref{fig3},
a non-vanishing $C$ can uplift (or pull down) the potential and realize a Minkowski vacuum. Finally, at the bottom of Fig.~\ref{fig3}, we show the values of $d^2V/d\theta^2$ at $\theta=0$ for the same parameter sets. We find that the $\theta$-direction is stable but these models are not suitable for PBH production.  

\end{appendix}

\bibliography{Bibliography}{}
\bibliographystyle{utphys}

\end{document}